\documentclass[aps,pra,preprint,groupedaddress,showpacs,amsmath,amssymb]{revtex4}

\usepackage{graphicx}
\usepackage{dcolumn}
\usepackage{bm}

\begin{document}

\preprint{2008.06.FanYang.Superradiance.v7}

\title{Resonant sequential scattering in two-frequency-pumping superradiance from a Bose-Einstein condensate}


\author{Fan Yang}
\author{Xiaoji Zhou}
\email[E-mail: ]{xjzhou@pku.edu.cn}
\author{Juntao Li}
\author{Yuankai Chen}
\author{Lin Xia}
\author{Xuzong Chen}
\email[E-mail: ]{xuzongchen@pku.edu.cn}
\affiliation{Institute of Quantum Electronics, School of Electronics
Engineering $\&$ Computer Science, Peking University, Beijing
100871, P. R. China}


\date{\today}

\begin{abstract}
We study sequential scattering in superradiance from a Bose-Einstein condensate pumped by a two-frequency laser beam. We find that the distribution of atomic side modes presents highly different patterns for various frequency difference between the two pump components. A novel distribution is observed, with a frequency difference of eight times the recoil frequency. These observations reveal that the frequency overlap between the end-fire modes related to different side modes plays an essential role in the dynamics of sequential superradiant scattering. The numerical results from a semiclassical model qualitatively agree with our observations.
\end{abstract}

\pacs{03.75.Kk, 42.50.Gy, 42.50.Ct, 32.80.Qk}

\maketitle

Due to its high degree of coherence, an atomic Bose-Einstein condensate (BEC) \emph{looks} different from thermal atomic clouds, in other words, the condensate scatters light in a coherent way. This property has been revealed in a series of experiments of superradiant light scattering from BECs \cite{Inouye:1999:sr, Schneble:2003:sr, Schneble:2004:srrm, Yoshikawa:2004:srrm, Fallani:2005:sr, Slama:2007:srcvt}, and applied to achieve phase-coherent amplification of matter-waves \cite{Inouye:1999:mwa, Kozuma:1999:mwa, Schneble:2004:srrm} and to probe the spatial coherence of ultracold gases \cite{Sadler:2007:srimg}. Several theoretical descriptions of these experiments have been published \cite{Moore:1999:srtheo, Mustecaplioglu:2000:srtheo, Pu:2003:srap, Cola:2004:srtheo, Zobay:2005:srtheo, Zobay:2006:srtheo, Uys:2007:srtheo}. In a typical BEC superradiance experiment, an elongated condensate is illuminated by an off-resonant pump laser along its short axis. A strong correlation between successive scattering events results in exponential growth of the scattered light and recoiling atoms. Due to the phase-matching effect and mode competition, the highly directional emissions of light travel along the long axis of the condensate, in the so-called end-fire modes \cite{Inouye:1999:sr}. Consequently, the recoiling atoms acquire well-defined momentum at $\pm 45^{\circ}$ angles with respect to the pump laser direction. These atomic modes are referred to as forward modes.

Meanwhile, atoms in the condensate may scatter a photon in the end-fire modes back into the pump mode and recoil at $\pm 135^{\circ}$ angles, forming the so-called backward modes. However, there is an energy mismatch of four times the recoil energy for this backward scattering, resulting from the increased kinetic energy of recoiling atoms. Considering the constraint of energy conservation for individual atom-photon scattering event, two regimes can be distinguished \cite{Schneble:2003:sr}. In the Kapitza-Dirac regime, the relevant time scale is so short compared to the recoil time that the energy mismatch is covered by a large energy uncertainty. Therefore, both forward and backward modes are populated. In the Bragg regime, by contrast, the pump pulse has a long duration, so backward scattering is suppressed and only forward modes are populated.

Compared to the superradiance first discussed by Dicke \cite{Dicke:1954:sr, Gross:1982:sr}, the superradiant light scattering from BECs is quite different that atoms may experience additional scattering cycles. The consequent sequential scattering transfers atoms to higher-order side modes successively and results in distribution of side modes, experimentally presented as atomic patterns in absorption images. A fan-shaped pattern was observed in the Bragg regime \cite{Inouye:1999:sr}, followed by an X-shaped one observed in the Kapitza-Dirac regime \cite{Schneble:2003:sr}, then they are regarded as characteristic signs of these two regimes, respectively. It's argued that the spatial propagation effects should be included to explain these distribution patterns \cite{Zobay:2005:srtheo, Zobay:2006:srtheo}.

Recently, to excite the backward scattering on a longer time scale, a two-frequency-pumping scheme was implemented \cite{BarGill:2007:srtf, vanderStam:2007:srtf}, where the pump beam consisted of two frequency components and the frequency difference between them was controlled to be around the energy mismatch. Within this new scheme, the enhancement to the backward scattering has been studied, however, sequential scattering and the consequent side-mode distribution were not discussed. We study the sequential scattering in this two-frequency-pumping scheme, and show that the distribution of side modes and the depletion behavior of the condensate are quite sensitive to the frequency difference.

\begin{figure}[tbp] 
   \centering
   \includegraphics[width=0.46\textwidth]{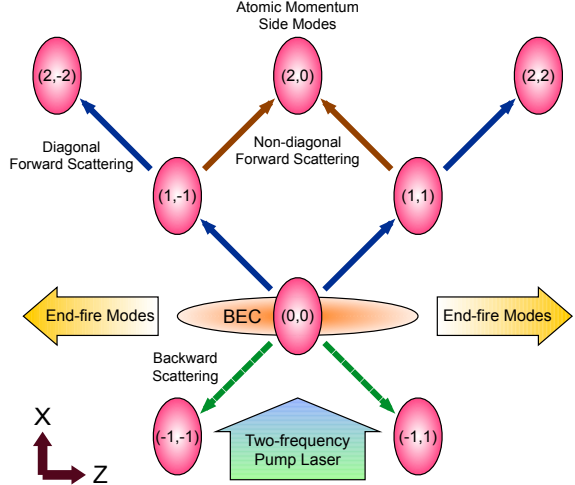} 
   \caption{\label{fig:ExpScheme}Schematic diagram of our experiment. The two-frequency pump beam is incident along the $x$ direction, with a linear polarization along the $y$ direction. The atomic side modes are denoted in momentum space, each labeled with a pair of integers which describe the order in the $x$ and $z$ directions, respectively. Within this notation, atoms of the condensate at rest are in mode $(0,0)$. A forward scattering event transfers an atom from mode $(n,m)$ to mode $(n+1,m\pm 1)$, and a backward event transfers one to mode $(n-1,m\pm 1)$.}
\end{figure}

In our experiment, a nearly pure Bose-Einstein condensate of about $2\times10^{5}$ $^{87}$Rb atoms in the $|F=2, m_{F}=2\rangle$ hyperfine ground state is generated in a quadrupole-Ioffe-configuration magnetic trap, with Thomas-Fermi radii of $50 \mathrm{\mu m}$ and $5 \mathrm{\mu m}$ along the axial and radial directions, respectively \cite{Ma:2006:Maj}. The laser beam from an external cavity diode laser is split into two equal-intensity beams. Their frequencies are shifted individually by acoustic-optical modulators (AOMs) which are driven by phase-locked radio frequency signals, therefore the frequency difference between the two beams can be controlled precisely. After that, two beams are combined to form our linear-polarized two-frequency pump beam, which is red detuned by $2\pi \times 2.3 \mathrm{GHz}$ from the $|F=2, m_{F}=2\rangle$ to $|F'=3, m_{F}'=3\rangle$ transition. In a typical experiment procedure, the pump pulse is incident along the short axis of the condensate, with its polarization perpendicular to the long axis (Fig.\ref{fig:ExpScheme}). This arrangement of polarization induces Rayleigh superradiance where all side modes possess the same atomic internal state \cite{Inouye:1999:sr}. The magnetic trap is shut off immediately after the pump pulse, then the distribution of atomic side modes is measured by absorption imaging after $21\mathrm{\mu s}$ of ballistic expansion.

\begin{figure}[tbp] 
   \centering
   \includegraphics[width=0.46\textwidth]{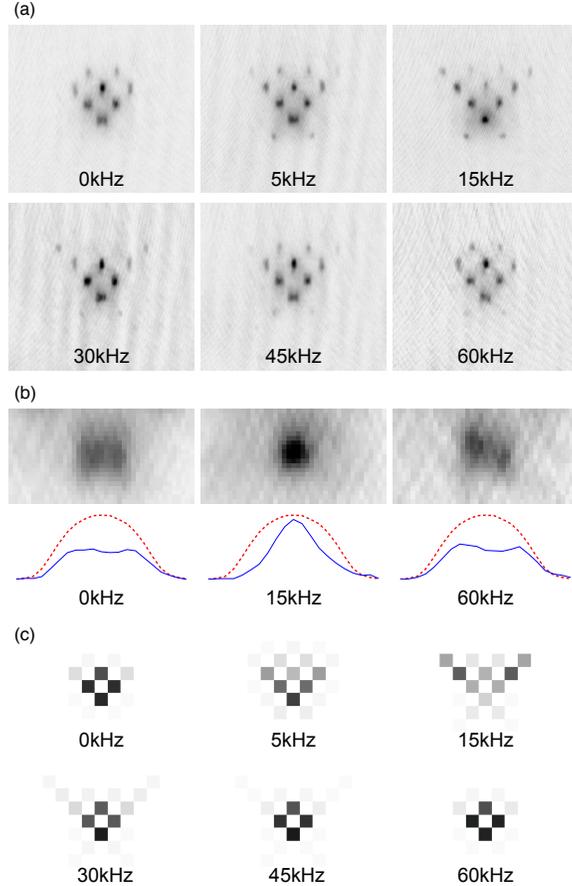} 
   \caption{\label{fig:Figure-Distri-FreqDiff}Distribution patterns of atomic side modes in two-frequency-pumping superradiance. (a) Experimental results with variable frequency difference $\Delta\omega/2\pi$. The pump pulse has a time-averaged central intensity of $150\mathrm{mW/cm^{2}}$ and a pulse duration of $200\mathrm{\mu s}$. The initial relative phase $\phi_{0}$ is controlled to be zero. The field of view of each image is $1.35\mathrm{mm} \times 1.21\mathrm{mm}$. (b) Magnification of the regions around mode $(0,0)$. Solid (blue) lines are profiles of the averaged optical density. The profile of an unperturbed condensate is plotted in dashed (red) lines for reference, with a vertical scale of $\times0.2$. (c) Calculated distribution patterns, with coupling constant $g = 1.25\times10^{6} \mathrm{s}^{-1}$.}
\end{figure}

The experiment is repeated with a variable frequency difference $\Delta \omega$. The pulse duration ($200\mathrm{\mu s}$) is long enough to exclude backward scattering using a single-frequency pump [see the $0\mathrm{kHz}$ image of Fig.\ref{fig:Figure-Distri-FreqDiff}(a)]. When $\Delta \omega$ is tuned close to the predicted two-photon resonance frequency of four times the recoil frequency ($4\omega_{r} = 2 \hbar k_{l}^{2}/M = 2\pi \times 15\mathrm{kHz}$), atoms in the backward modes are clearly observed, as reported in Ref.\cite{BarGill:2007:srtf}. When $\Delta \omega$ is tuned far from the resonance frequency, by contrast, atoms in the backward modes are negligible [see the $45\mathrm{kHz}$ and $60\mathrm{kHz}$ images of Fig.\ref{fig:Figure-Distri-FreqDiff}(a)]. We note that the total number of scattered atoms also shows sensitivity to the initial relative phase $\phi_{0}$ between the two pump components, which may be ascribed to the nonlinearity of the self-amplification process in BEC superradiance. A fixed phase of $\phi_{0}=0$ is used throughout this paper.

On the other hand, by merely controlling the frequency difference, we obtain highly different distribution patterns [Fig.\ref{fig:Figure-Distri-FreqDiff}(a)]. When $\Delta\omega$ is tuned to $2\pi \times 15\mathrm{kHz}$, i.e., the resonance frequency for exciting the backward scattering, diagonal modes ($|n| = |m|$) are densely populated and an X-shaped-like pattern emerges, even though our pump duration goes beyond the bounds of the Kapitza-Dirac regime (several tens of $\mathrm{\mu s}$ for Rubidium). However, the population is obviously unbalanced for the forward and backward modes, which is different from the almost symmetrical population obtained in the Kapitza-Dirac regime (see Fig.1 in Ref.\cite{Schneble:2003:sr}), suggesting a limited efficiency of exciting backward scattering using a two-frequency pump. This forward-backward asymmetry also suggests that the proposed atom-pair production \cite{Pu:2003:srap, vanderStam:2007:srtf} is not responsible for the diagonal tendency of forward modes in this situation. As $\Delta\omega$ is tuned away from $2\pi \times 15\mathrm{kHz}$, atoms start to cluster in the non-diagonal modes, especially in mode $(2,0)$. Eventually, when $\Delta\omega$ is larger than $2\pi\times45\mathrm{kHz}$, the side-mode distribution transforms back to the fan-shaped pattern, exactly the same to the pattern obtained in the Bragg regime of single-frequency pumping.

The depletion behavior of the condensate also shows sensitivity to the frequency difference. In an experiment of Rayleigh superradiance using a single-frequency pump, depletion in the center of the condensate was reported for the Bragg regime \cite{Schneble:2003:sr}. In our two-frequency experiment, similar central depletion is observed with off-resonant $\Delta\omega$ [see the 0kHz and 60kHz images of Fig.\ref{fig:Figure-Distri-FreqDiff}(b)]. It's the opposite result for $\Delta\omega=2\pi\times15\mathrm{kHz}$ that the condensate has large depletion at its two tips [see the 15kHz image of Fig.\ref{fig:Figure-Distri-FreqDiff}(b)]. A semiclassical model of single-frequency-pumping superradiance proposed a correlation between the central depletion and the onset of population growth in mode $(2,0) $\cite{Zobay:2006:srtheo}. Indeed, we never observe central depletion when the pump intensity is reduced that only first-order side modes get populated. Despite the difference  of pump composition between our experiment and the theoretical model, our observations suggest the same correlation. We note that the spatial inhomogeneities in BEC superradiance have been studied experimentally in Ref.\cite{Sadler:2007:srimg}. However, it's Raman superradiance excited in that experiment where only first-order side modes existed \cite{Schneble:2004:srrm, Yoshikawa:2004:srrm}. Higher absorption of the pump beam, which indicates higher depletion of atoms, was found at two tips of the condensate. A numerical simulation qualitatively agrees with their results \cite{Uys:2007:srtheo}. To our knowledge, so far, no central depletion has been reported for Raman superradiance.

\begin{figure}[tbp] 
   \centering
   \includegraphics[width=0.46\textwidth]{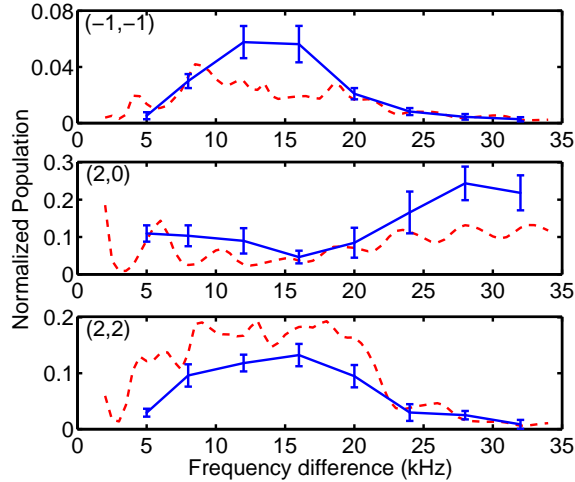} 
   \caption{\label{fig:spectrum}The normalized populations of modes $(-1,-1)$, $(2,0)$ and $(2,2)$, as functions of the frequency difference. Solid (blue) lines connect experimental results, which are normalized by the measured total atom number in all side modes. The time-averaged pump intensity is  $125\mathrm{mW/cm^{2}}$, which is slightly reduced to keep negligible atoms in higher-order side modes. Dashed (red) lines are calculated results with coupling constant $g = 1.05\times10^{6} \mathrm{s}^{-1}$, normalized by a fixed atom number of $2\times10^{5}$.}
\end{figure}

The spectroscopic responses of modes $(-1,-1)$, $(2,0)$ and $(2,2)$ are plotted in Fig.\ref{fig:spectrum}. These three modes are selected to represent the backward, the non-diagonal forward and the diagonal forward scattering, respectively. As reported in Ref. \cite{BarGill:2007:srtf}, the spectrum of mode $(-1,-1)$ shows broad resonance centered around the predicted $4\omega_{r}/2\pi \approx 15\mathrm{kHz}$. Meanwhile, the spectra of forward modes $(2,0)$ and $(2,2)$ also clearly show broad resonance centered around $15\mathrm{kHz}$. However, these two second-order modes show opposite responses, indicating a competition between the non-diagonal and the diagonal forward scattering.

\begin{figure}[tbp] 
   \centering
   \includegraphics[width=0.46\textwidth]{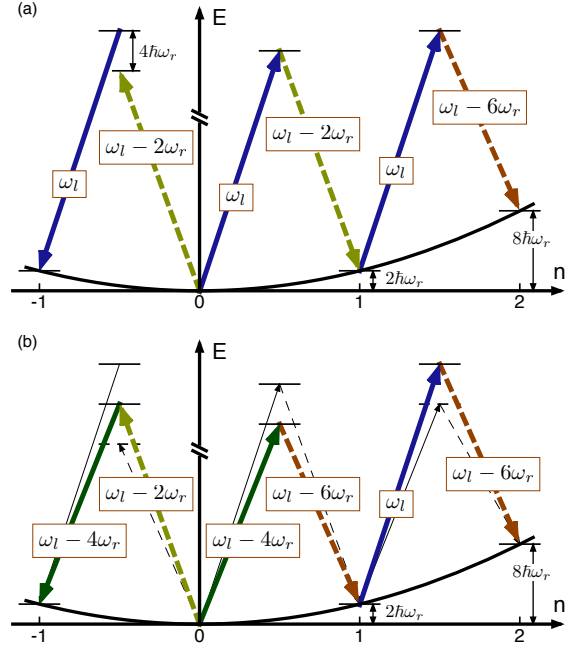} 
   \caption{Energy levels of the diagonal ($|n| = |m|$) side modes, along with the relevant pump (solid arrows) and end-fire modes (dashed arrows). In our experiment, all atoms are in the same internal state, thus energy levels are determined by the kinetic energy, $E_{n} \approx n^{2} \hbar^{2} k_{l}^{2}/M = 2n^{2}\hbar\omega_{r}$. (a) In the case of single-frequency pumping, the first-order superradiance between modes $(0,0)$ and $(1,\pm1)$ is intrigued by a spontaneous scattering, and the first-order end-fire mode of $\omega_{l}-2\omega_{r}$ forms in a self-amplified manner. However, due to the parabolically increased kinetic energy of atomic side modes, the frequency of the second-order end-fire mode is  $4\omega_{r}$ lower than that of the first-order mode, so the second-order scattering can't be excited resonantly by existing light fields. Meanwhile, the first-order backward scattering is off-resonant of $4\omega_{r}$ and suppressed in the Bragg regime. (b) In the case of two-frequency pumping with a frequency difference of $\Delta\omega = 4\omega_{r}$, the first-order backward scattering can be excited resonantly. Meanwhile, the first- and second-order end-fire modes share a common frequency component of $\omega_{l}-6\omega_{r}$, resulting in a stimulating interplay between the first- and second-order diagonal scattering. }
   \label{fig:Figure-Single-Two-Freq}
\end{figure}

The emergence of the diagonal distribution with a $\Delta\omega$ tuned close to $4\omega_{r}$ can be understood by analyzing frequency composition of the end-fire modes. In a forward-scattering event, the scattered photon has a lower frequency than that in the pump mode, compensating for the increased kinetic energy of the recoiling atom. In the laboratory frame, the momentum of mode $(n,m)$ is $\sqrt{n^{2} \hbar^{2} k_{l}^{2} + m^{2} \hbar^{2} k^{2}} \approx \sqrt{n^{2}+m^{2}}\hbar k_{l}$, and the kinetic energy is $(n^{2}+m^{2})\hbar^{2}k_{l}^{2}/2M = (n^{2}+m^{2})\hbar\omega_{r}$, where $k_{l}$ and $k$ are wave numbers of the pump and end-fire modes, respectively. For a diagonal forward-scattering event between modes $(n, n)$ and $(n+1, n+1)$, the relevant $(n+1)$th-order end-fire mode has a frequency decrease of $2(2n+1)\omega_{r}$, resulting a ``detuning barrier'' of $4\omega_{r}$ between neighboring order side modes \cite{Zobay:2007:srtheo}. Consequently, in the Bragg regime of single-frequency pumping, the existing light fields can't excite higher-order diagonal scattering resonantly [Fig.\ref{fig:Figure-Single-Two-Freq}(a)]. For two-frequency pumping, a lower-frequency pump component is introduced to compensate for the energy mismatch of backward scattering. Meanwhile, it compensates for the detuning barrier between the diagonal forward modes. Thus diagonal sequential scattering can be excited resonantly, and diagonal side modes get populated quickly [Fig.\ref{fig:Figure-Single-Two-Freq}(b)]. We note that the non-diagonal forward scattering is also on-resonant, but it's suppressed in the competition with the diagonal scattering, due to the poor spatial overlap between atomic modes and relevant light fields, as discussed in Ref.\cite{Zobay:2006:srtheo}.

\begin{figure}[tbp] 
   \centering
   \includegraphics[width=0.46\textwidth]{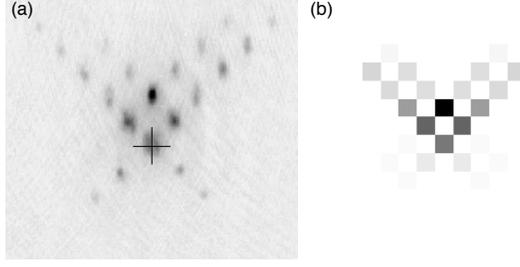} 
   \caption{\label{fig:Figure-Distri-30kHz}A new side-mode distribution for a frequency difference of eight times the recoil frequency. (a) The experiment result of $\Delta\omega = 2\pi\times30\mathrm{kHz}$, with a time-averaged pump intensity of $I=170\mathrm{mw/cm^{2}}$. The cross in the image indicates the center of mode $(0,0)$. (b) A corresponding numerical pattern with $g = 1.5\times10^{6} \mathrm{s}^{-1}$.}
\end{figure}

Interestingly, when $\Delta\omega$ is tuned around $2\pi\times30\mathrm{kHz}$, a novel distribution pattern emerges, which has never been observed with a single-frequency pump [see Fig.\ref{fig:Figure-Distri-30kHz}(a), or the $30\mathrm{kHz}$ image in Fig.\ref{fig:Figure-Distri-FreqDiff}(a)]. In this distribution, non-diagonal mode $(2,0)$ is highly populated, but combined with an obvious tendency toward the diagonal population. This distribution can be understood that $\Delta\omega=2\pi\times30\mathrm{kHz}=8\omega_{r}$ exactly fills the frequency difference between the first- and third-order end-fire modes, and results in an alternate stimulation in diagonal sequential scattering. However, this alternate effect is weaker compared to that of $\Delta\omega=4\omega_{r}$, so the stimulation chain along diagonal directions needs more time to be established, and the resonant non-diagonal scattering dominates in the early stage. So far, we have not observed any similar distribution with a frequency difference of $12\omega_{r}$ or $16\omega_{r}$.

By including spatial propagation effects, an one-dimensional semiclassical model developed by Zobay \emph{et al.} reproduced some characteristic properties of BEC superradiance of single-frequency pumping \cite{Zobay:2005:srtheo,Zobay:2006:srtheo}. It's generalized to accommodate to a pump laser consisting of two frequency components, $\omega_{l}= c k_{l}$ and $\omega_{l}-\Delta\omega$ \cite{BarGill:2007:srtf}. These two components are supposed to have the same intensity and polarization here. Since $\Delta\omega \ll \omega_{l}$, the difference between their wave vectors is omitted, then the coupling constant (13) in Ref.\cite{Zobay:2006:srtheo} is generalized to
\begin{equation}
	\bar{g}(t) = g \left(1 + e^{i (\Delta\omega t + \phi_{0})} \right),
\end{equation}
where $g=\sqrt{3 \pi c^{3} R / (2 \omega_{l}^{2} A L)}$, and $R$ the Rayleigh scattering rate of either pump component, $A$ the average cross-sectional area of the BEC perpendicular to its long axis, $L$ the length of the BEC. The initial relative phase $\phi_{0}$ is always zero in our discussion.

With the dimensionless coordinates $\tau = 2\omega_{r}t$ and $\xi = k_{l}z$, the coupled evolution equations of atomic side modes are given by
\begin{eqnarray}
	i \frac{\partial\psi_{nm}(\xi,\tau)}{\partial \tau} = && - \frac{1}{2}\frac{\partial^{2}\psi_{nm}(\xi,\tau)}{\partial \xi^{2}} - i m \frac{\partial\psi_{nm}(\xi,\tau)}{\partial \xi} \nonumber\\
	&&+\kappa^{*}(\tau) e_{+} \psi_{n+1,m-1} e^{-i(n-m)\tau} \nonumber\\
	&&+ \kappa^{*}(\tau) e_{-} \psi_{n+1,m+1} e^{-i(n+m)\tau} \nonumber\\
	&&+\kappa(\tau) e_{+}^{*} \psi_{n-1,m+1} e^{i(n-m-2)\tau} \nonumber\\
	&&+\kappa(\tau) e_{-}^{*} \psi_{n-1,m-1} e^{i(n+m-2)\tau},
\end{eqnarray}
and the equations for optical fields read
\begin{eqnarray}
	&&e_{\pm}(\xi,\tau) = \mp i \frac{\kappa(\tau)}{\chi} \int_{\mp\infty}^{\xi}\mathrm{d}\xi' \nonumber\\
	&&\times \sum_{n,m} e^{i(n \mp m)\tau} \psi_{n,m}(\xi', \tau) \psi_{n+1,m \mp 1}^{*}(\xi', \tau),
\end{eqnarray}
with $\kappa(\tau) = \bar{g}(\tau)\sqrt{k_{l} L}/2\omega_{r}$ and $\chi = c k_{l}/2\omega_{r}$.

The calculated distribution patterns with an initial seed of one atom in modes $(1,1)$ and $(1,-1)$ are shown in Fig.\ref{fig:Figure-Distri-FreqDiff}(c), and reproduce the transformation observed in our experiment. The calculated spectroscopic responses are plotted in Fig.\ref{fig:spectrum}, which also present similar resonance behavior. These numerical results qualitatively agree with our experimental results, suggesting that this generalized model is valid for a two-frequency pump.

\begin{figure}[tbp] 
   \centering
   \includegraphics[width=0.46\textwidth]{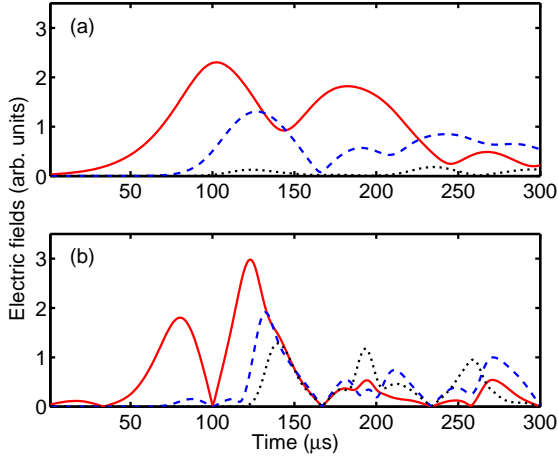} 
   \caption{\label{fig:Figure-end_fire_modes}Moduli of the end-fire mode components $e_{+}^{(0,0)}$ (solid line), $e_{+}^{(1,1)}$ (dashed line) and $e_{+}^{(2,2)}$ (dotted line). (a) Single-frequency pumping. $\Delta\omega = 0$, $g = 0.85\times10^{6} \mathrm{s}^{-1}$. (b) Two-frequency pumping. $\Delta\omega = 2\pi\times 15\mathrm{kHz}$, $g = 1.25\times10^{6} \mathrm{s}^{-1}$. }
\end{figure}

Following the treatment in Sec.\uppercase\expandafter{\romannumeral5} of Ref.\cite{Zobay:2006:srtheo}, the end-fire mode is decomposed as $e_{+} = \sum_{n,m} e_{+}^{(n,m)}$, where each term $e_{+}^{(n,m)}$ is interpreted as the contribution to the total field arising from scattering between the side modes $(n,m)$ and $(n+1,m-1)$. The calculated time dynamics of $e_{+}^{(0,0)}$, $e_{+}^{(1,1)}$ and $e_{+}^{(2,2)}$, i.e., the first-, second- and third-order diagonal components of $e_{+}$, are plotted in Fig.\ref{fig:Figure-end_fire_modes}. We find that, in the case of single-frequency pumping, $e_{+}^{(1,1)}$ grows gradually, and $e_{+}^{(2,2)}$ retains small amplitude throughout the time scale of calculation [Fig.\ref{fig:Figure-end_fire_modes}(a)]. However, in the case of two-frequency pumping with $\Delta\omega = 2\pi \times 15\mathrm{kHz}$, a sudden rise of $e_{+}^{(1,1)}$ and $e_{+}^{(2,2)}$ appears around $120\mathrm{\mu s}$ [Fig.\ref{fig:Figure-end_fire_modes}(b)], supporting our analysis of the stimulating interplay among diagonal modes.

In conclusion, the sequential scattering in BEC superradiance with a two-frequency pump is studied. The distribution of atomic side modes, which reflects the dynamics of sequential scattering directly, shows obvious sensitivity to the frequency difference $\Delta\omega$ between the two pump components. When $\Delta\omega$ is tuned to the predicted resonance frequency of $4\omega_{r}$, an almost diagonal distribution is obtained with a long pulse duration of $200\mathrm{\mu s}$, different from the case of single-frequency pumping where x-shaped pattern is only observed with short and intense pump pulse. A novel distribution is obtained with a frequency difference of $8\omega_{r}$. We ascribe these phenomena to that diagonal sequential scattering can be excited resonantly when $\Delta\omega$ is tuned to match the detuning barrier between atomic side modes, successively or alternately. Furthermore, the depletion behavior of the condensate also shows sensitivity to $\Delta\omega$, and our experiment results support the expectation that the central depletion correlates with the onset of mode $(2,0)$. The numerical simulations from a generalized one-dimensional semiclassical model qualitatively agree with our observations, and support above analysis. Our results present the ability to control the sequential dynamics of BEC superradiance on a large time scale by composing the pump beam. We expect richer possibilities for more pump components or a pulsed pump.


\begin{acknowledgments}
We would like to thank O. Zobay for his help on numerical simulation. This work is partially supported by the state Key Development Program for Basic Research of China (No.~2005CB724503, 2006CB921401 and 2006CB921402), NSFC(No. 60490280 and 10574005).
\end{acknowledgments}

\bibliography{200805_YF_SR}

\end{document}